\documentclass[structabstract]{aa}
\usepackage{graphicx} 
\usepackage{txfonts} 
\usepackage{longtable}
\usepackage{float}

\begin{document}

\title{The Effects of Variations in Nuclear Interactions on Nucleosynthesis in Thermonuclear Supernovae}

\author{Anuj Parikh \inst{1,2}
   \and Jordi Jos\'e \inst{1,2}
   \and Ivo R. Seitenzahl \inst{3,4}
   \and  Friedrich K. R\"opke \inst{3}
   }

\offprints{A. Parikh}

\institute{Departament de F\'isica i Enginyeria Nuclear, EUETIB,
            Universitat Polit\`ecnica de Catalunya --- BarcelonaTech, 
            c/Comte d'Urgell 187, 
            E-08036 Barcelona, 
            Spain,\
\email{anuj.r.parikh@upc.edu}
            \and 
            Institut d'Estudis Espacials de Catalunya, 
            c/Gran Capit\`a 2-4, 
            Ed. Nexus-201, 
            E-08034 Barcelona, 
            Spain\
            \and
            Universit\"at W\"urzburg, Emil-Fischer-Strasse 31, D-97074 W\"urzburg, Germany\
            \and
            Max-Planck-Institut f\"ur Astrophysik, Karl-Schwarzschild-Strasse 1, D-85748 Garching, Germany\
      }
       
\date{\today}

\abstract{Type Ia supernova explosions are violent stellar events important for their contribution to the cosmic abundance of iron peak elements and for their role as cosmological distance indicators.}
         {The impact of nuclear physics uncertainties on nucleosynthesis in thermonuclear supernovae has not been fully explored using comprehensive and systematic studies with multiple models.  To better constrain predictions of yields from these phenomena, we investigate thermonuclear reaction rates and weak interaction rates that significantly affect yields in our underlying models.}
         {We have performed a sensitivity study by post-processing thermodynamic histories from two different hydrodynamic, Chandrasekhar-mass explosion models.  We have individually varied all input reaction and, for the first time, weak interaction rates by a factor of ten (up and down) and compared the yields in each case to yields using standard rates.}
         {Of the 2305 nuclear reactions in our network, we find that the rates of only 53 reactions affect the yield of any species with an abundance of at least 10$\mathrm{^{-8} M_{\sun}}$ by at least a factor of two, in either model.   The rates of the $^{12}$C($\alpha,\gamma$), $^{12}$C+$^{12}$C, $^{20}$Ne($\alpha,p$), $^{20}$Ne($\alpha,\gamma$) and $^{30}$Si($p,\gamma$) reactions are among those that modify the most yields when varied by a factor of ten.  From the individual variation of 658 weak interaction rates in our network by a factor of ten, only the stellar $^{28}$Si($\beta^{+}$)$^{28}$Al, $^{32}$S($\beta^{+}$)$^{32}$P and $^{36}$Ar($\beta^{+}$)$^{36}$Cl rates significantly affect the yields of species in a model.  Additional tests reveal that reaction rate changes over temperatures $T > 1.5$ GK have the greatest impact, and that ratios of radionuclides that may be used as explosion diagnostics change by a factor of $\lesssim2$ from the variation of individual rates by a factor of 10.}
         {Nucleosynthesis in the two adopted models is relatively robust to variations in individual nuclear reaction and weak interaction rates.  Laboratory measurements of a limited number of reactions would, however, help to further constrain model predictions.  As well, we confirm the need for a detailed, consistent treatment for all relevant stellar weak interaction rates since simultaneous variation of these rates (as opposed to individual variation) has a significant effect on yields in our models.}


\titlerunning{Effects of Variations in Nuclear Interactions on in Thermonuclear Supernovae} 
\authorrunning{A. Parikh et al.} 

\maketitle

\section{Introduction}

Type Ia supernovae (hereafter, SNe~Ia) have become valuable cosmological tools. Through calibrated 
light curve analysis, they have been used as probes to outline the geometrical structure of the 
Universe, unraveling  its unexpected acceleration stage (Riess et al. 1998; Schmidt et al. 1998; Perlmutter et al. 1999; 
for constraints on dark energy and on the cosmic expansion 
history, see recent work by Astier et al. 2011; Blake et al. 2011; Suzuki et al. 2012; and 
references therein). 

SNe~Ia are spectroscopically classified by the absence of hydrogen (Balmer)
emission lines and the presence of a prominent P-Cygni absorption feature near 6150 $\AA$ due to blueshifted 
Si II (Wheeler \& Harkness 1990; Filippenko 1997)\footnote{Some SNe~Ia are, however, anomalous in this regard. SN 2002ic (Hamuy et al. 2003), for instance,
while exhibiting other standard features common to all SNe~Ia, unequivocally showed
broad-line H$_\alpha$ emission. This has been interpreted as
proof of a SN~Ia interacting with H-rich circumstellar material. Other SN
2002ic-like events include SN 2005gj, PTF 11kx (Dilday et al. 2012) and SN 2008J (Taddia et al. 2012).}. 
Whereas the first observational constraint places limits on the maximum 
amount of hydrogen that can be present in the expanding atmosphere of the star (i.e., $M_{\rm H} \leq 
0.03-0.1 \mathrm{M}_{\sun}$), the second feature suggests that the 
outermost ejected shells contain intermediate-mass
elements from nuclear processing.

The increasing number of supernovae discovered has revealed some diversity among SNe~Ia, raising doubts 
upon the historically postulated uniqueness of the progenitor system. 
Already two decades ago estimates had indicated that only about 85\% of the observed SNe~Ia belong to a
homogeneous class of events (Branch et al. 1993), 
with a dispersion of only $\Delta M \leq 0.3$ $mag$ when normalized to peak luminosity
(see Cadonau et al. 1985; Hamuy et al. 1996). These SNe~Ia are known as ``Branch-normals", 
with canonical examples such as SNe 1972E, 1981B, 1989B or 1994D. 
A more recent classification of SNe~Ia (Li et al. 2011, in a volume-limited sample) found that 
the number of SNe~Ia deviating from this homogeneous class of objects is closer to $\sim$30\% 
(see also Li et al. 2000; Branch 2004; Kasen et al. 2009). Within the minority group,
likely progenitor systems and explosion models have been identified recently for two sub-classes.
For SN 2002cx-like SNe, observable features are well explained by weak deflagrations in near-Chandrasekhar-mass carbon--oxygen (CO) white dwarf stars  (WDs) leaving bound remnants (Jordan et al. 2012; Kromer et al. 2013); for SN 1991bg-like SNe, the peculiar spectra, colours, and low expansion velocities of this sub-luminous class are reproduced by mergers of two, relatively light CO WDs of nearly equal mass (Pakmor et al. 2010). 

The most promising progenitor scenarios that have been proposed for spectroscopically
``normal" SNe~Ia (e.g., Livio 2000; Hillebrandt \& Niemeyer 2000; Isern et al. 2011; Howell 2011; Hillebrandt et al. 2013) include: the single degenerate scenario, consisting of a non-degenerate companion star that transfers hydrogen-rich (or, possibly, helium-rich) matter onto a CO WD (Whelan \& Iben 1973); the double degenerate scenario, consisting of two merging CO WDs, such that the total mass exceeds the Chandrasekhar limit (Iben \& Tututkov 1984; Webbink 1984); and, a scenario where a sub-Chandrasekhar-mass CO WD accretes stably from a companion star and explodes before reaching the Chandrasekhar limit (Taam 1980; Iben et al. 1987).  Other scenarios may indeed be possible.  In spite of the large number of published SN~Ia explosion models for proposed scenarios, detailed predictions of the associated nucleosynthesis have been calculated and published only for a very limited subset of near-Chandrasekhar-mass WD models in the single degenerate scenario. To the best of our knowledge, 1D model yields exist in the literature only for variants of pure deflagrations (Thielemann et al. 1986; Nomoto et al. 1997; Woosley 1997; Iwamoto et al. 1999; Maeda et al. 2010); 2D model yields have been published for gravitationally confined detonation (Meakin et al. 2009) and delayed-detonation models (Maeda et al. 2010); and detailed tables of yields from 3D models are available only for pure deflagrations (Travaglio et al. 2004; R\"opke et al. 2006) and a suite of 14 delayed-detonation models (Seitenzahl et al. 2013).

These models have revealed that the nucleosynthesis in SNe~Ia depends critically on the peak temperature
achieved and the density at which the thermonuclear runaway occurs.  As well, the specific composition of the WD (in particular the amount and distribution of $^{12}$C and $^{22}$Ne) plays a central role (see e.g., Chamulak et al. 2007, 2008; Townsley et al. 2009, and references therein), as it influences properties such as the ignition density, the release of energy, the flame speed and the specific density at which the initial deflagration may transform into a detonation.  In general, the abundance pattern of the ejecta is the result of five burning regimes: ``normal" and ``$\alpha$-rich" freeze-out from nuclear statistical equilibrium (NSE)
in the inner regions, and incomplete Si-, O-, and C/Ne-burning in the outermost layers (Thielemann et al. 1986; Woosley 1986).  SNe~Ia are galactic factories of $^{56}$Fe, producing about half of the iron content in the Milky Way
(Acharova et al. 2012) and perhaps 65 -- 70\% of the Fe in the solar neighbourhood (Mennekens et al. 2013). 
Hence, reproducing the chemical abundance pattern around the Fe-peak is a critical test in SN~Ia 
modeling.  For decades, models systematically overproduced neutron-rich species 
such as $^{54}$Cr or $^{50}$Ti with respect to Solar System values (Woosley 1990; 
Thielemann et al. 1997; Nomoto et al. 1997).
The agreement improved following the revision of key stellar weak interaction rates 
(Langanke \& Martinez--Pinedo 2000) and through the use of more recent 3D explosion models (Seitenzahl et al. 2013).  These results emphasize the critical roles played by both the nuclear physics input and the modeling techniques employed.

Nonetheless, in spite of their demonstrated importance, the impact of nuclear physics uncertainties on the nucleosynthesis produced in SNe~Ia has not been analyzed at the same level of detail as for other astrophysical scenarios, such as classical nova explosions or type I X-ray bursts.  Most efforts have focused on determining the role of the $^{12}$C+$^{12}$C reaction (e.g., Spillane et al. 2007; Bravo et al. 2011), since it triggers the explosion when the temperature exceeds $\approx700$ MK.  Progress has been reported in very recent work by Bravo \& Mart\'inez-Pinedo (2012), in which thermonuclear reaction rates of importance were investigated using a single 1D delayed-detonation model of a Chandrasekhar-mass WD.

The goal of this paper is to investigate the sensitivity of nucleosynthesis in thermonuclear supernovae to variations of both the nuclear reaction and weak interaction rates involved during the explosion. 
Since all currently available yield predictions are for near Chandrasekhar-mass models, we have chosen for our study two representative cases in this class: the W7 model (Nomoto et al. 1984), which has been serving as the fiducial SN~Ia explosion model in the community for almost three decades, and a standard 2D delayed-detonation (DDT) model.  Since our method involves recalculating the nucleosynthesis in each model every time a reaction or weak interaction rate is varied (resulting in over several million individual post-processing calculations), performing our study for a large suite of explosion models is not yet feasible.

In the following, we begin by briefly describing the two underlying models employed for our studies.  Next, we present and compare results obtained from individually varying each of the rates in our nucleosynthesis network to assess its impact on SNe~Ia yields, for each of our two models. Finally, we discuss some additional tests we have performed to motivate new experimental measurements and to link our results to SN~Ia observables. 

\section{Explosion models}
\label{models}

For our sensitivity studies we have chosen to
post-process the results from two SN~Ia explosion models: a generic two dimensional Chandrasekhar-mass delayed-detonation model (similar
to models from Kasen et al. 2009), and the fiducial
W7 pure deflagaration model (Nomoto et al. 1984).
The $^{56}$Ni mass ejected is 0.66 $\mathrm{M_{\sun}}$ in the W7 model and 0.68 $\mathrm{M_{\sun}}$ in the DDT model; these values are rather typical for normal SN~Ia (see, e.g. Stritzinger
et al. 2006). The details of the models are not the focus
here, they merely serve to define a set of reasonable thermodynamic
conditions that arise in thermonuclear supernovae
for the purpose of examining the sensitivity of SN~Ia yields to variations of the input reaction and weak interaction rates.  

\subsection{Delayed-detonation model}

\label{DDTmod}

The delayed-detonation model employed was a two dimensional
axisymmetric hydrodynamic explosion
of a 1.40 $\mathrm{M_{\sun}}$ cold WD with a central density of $2.9 \times 10^{9}$ g cm$^{-3}$.   The initial chemical composition was 47.5\% $^{12}$C, 50\% $^{16}$O, and 2.5\% $^{22}$Ne homogeneously
distributed throughout the WD. The $^{22}$Ne content parametrizes the neutron excess and corresponds
to an electron fraction of $Y_{e}$ = 0.49886. The simulation was performed
with the \textsc{LEAFS} code, which integrates
the discretized reactive Euler equations with a finite volume
method. The hydrodynamics solver is essentially the
\textsc{PROMETHEUS} implementation (Fryxell et al. 1989) of the ``piecewise parabolic method" by Colella \& Woodward
(1984). 

Subsonic deflagration flames and supersonic
detonation fronts were modeled as level set discontinuities between
nuclear fuel and nuclear ash (Osher \& Sethian 1988; Smiljanovski et al. 1997; Reinecke et al. 1999).
Any material traversed by these fronts was burned to a
composition that depends on fuel density and the mode of
burning (deflagration or detonation). The corresponding energy
was then immediately released behind the front. For detonations,
we have used the energy release data from the tables of
Fink et al. (2010). The speed of the detonations was modeled
as in Fink et al. (2010): at high densities ($\rho > 10^{7}$ g cm$^{-3}$), speeds were taken from Gamezo et al.
(1999); at low densities, Chapman--Jouguet like speeds were
calculated for the incomplete burning yields in our detonation
tables. For deflagrations, we have used the same energy
release table that was employed by Seitenzahl et al. (2011).
Since in two dimensions our usual subgrid-turbulence
based approach to model the deflagration-to-detonation
transition (DDT) probabilities is not applicable, we have modeled
the DDT using the method of Kasen
et al. (2009). For the critical Karlovitz number we have chosen 
$\mathit{Ka}_\mathrm{crit} = 250$, and we have limited the density at
which a DDT may occur to the interval $0.6 < \rho$ / 10$^{7}$ g cm$^{-3}$ $<
1.2$.

\begin{figure}  
\begin{center}  
\includegraphics[scale=0.9]{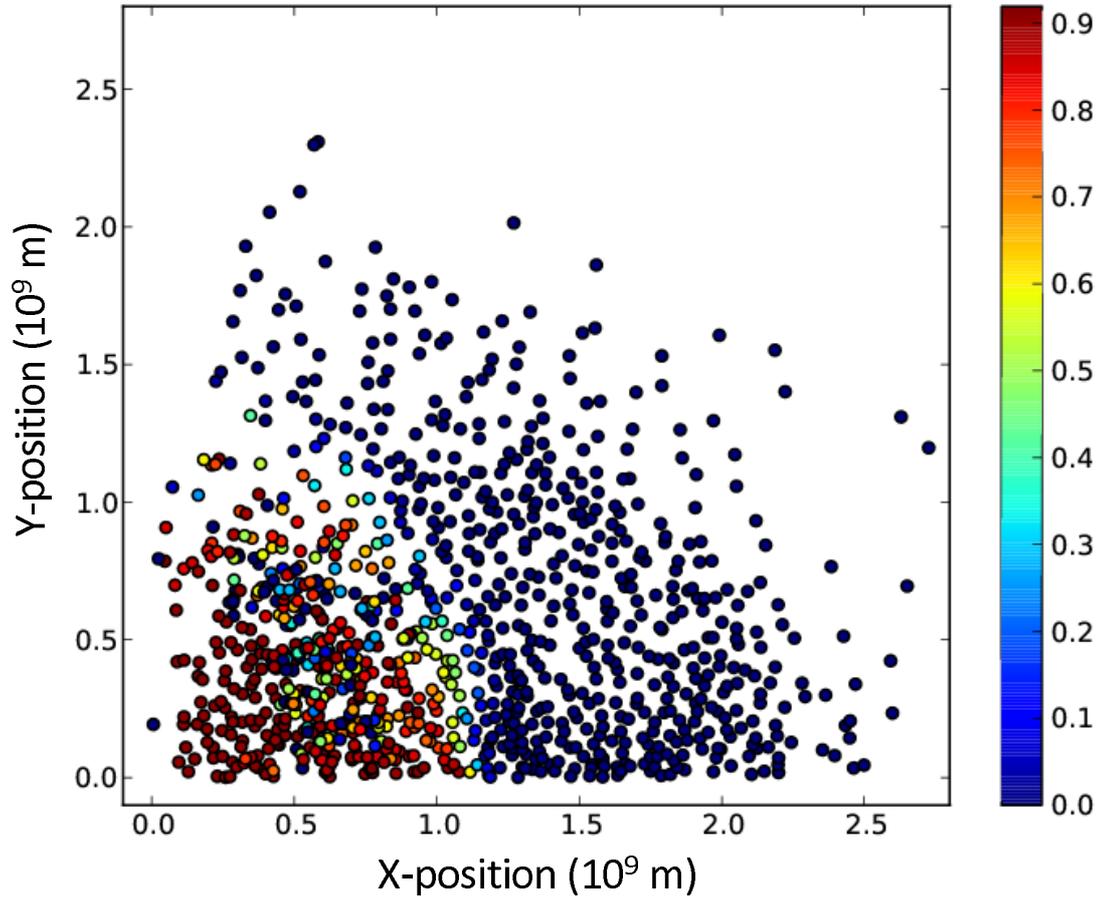}
\caption{Tracer particle positions for the adopted delayed-detonation model at $t = 100$s from first ignition of the deflagaration.  The tracer particles are coloured (online) according to
the mass fraction of $^{56}$Ni present.}
\label{fig_DDT_56Ni}  
\end{center}  
\end{figure}

The simulation method is similar to that employed in Kasen
et al. (2009), except that here we have assumed reflectional symmetry across the equator as well.  This is to facilitate analysis of the nucleosynthetic yields of the model using the tracer particle method (Travaglio et al. 2004), in which the temperature-density-time profile of each tracer particle is recorded and later used within post-processing calculations.  With the above assumption we attain twice the spatial sampling
of the explosion ejecta with the same number of tracer
particles.  We have used the
variable mass tracer particle method of Seitenzahl et al. (2010), which allows us to better resolve the lower density regions
of incomplete burning without greatly increasing the total number of tracer particles. The resolution was
512 $\times$ 512 computational cells and 1010 tracer particles were distributed to sample the mass distribution of the star.  This approach adequately samples the different nucleosynthesis regimes in a multipoint ignition delayed-detonation model and provides yields that are accurate to a few percent or better for the more abundant nuclides (Seitenzahl et al. 2010).  To illustrate, Figure~\ref{fig_DDT_56Ni} shows tracer particle positions for this model at $t = 100$s from the first ignition of the deflagaration.

\subsection{W7 model}
\label{W7mod}

The W7 model of Nomoto et al. (1984) is commonly used as a reference when general features of SN~Ia are discussed.  W7 is a one dimensional, fast deflagration explosion model of a $1.38$ $\mathrm{M_{\sun}}$
WD consisting of 50\% $^{16}$O, 47.5\% $^{12}$C, and 2.5\%
$^{22}$Ne by mass, homogeneously distributed. To model the acceleration of the convectively-driven deflagration
wave in their hydrodynamics code,
Nomoto et al. (1984) used the time-dependent mixing length theory of Unno (1967). They chose $\alpha=0.7$ for the
mixing length $l=\alpha H_p$, where $H_p$ is the pressure scale height. In their hydrodynamical
simulation only an $\alpha$-chain network was included to model the nuclear
energy release. Later, Thielemann et al. (1986) calculated the detailed nucleosynthesis 
for the W7 model by post-processing the thermodynamic histories of
the 172 zones with a reaction network comprising 259 nuclear species. 
The strong overproduction of neutron-rich Fe-group nuclei initially noted from this model was later attributed to the relatively large Fe-group electron capture rates
of Fuller et al. (1982) that had been used (see Iwamoto et al. 1999). Use of newer, reduced Fe-group 
electron capture rates (Langanke \& Mart\'inez-Pinedo 2000) improved the agreement between W7 model yields and solar system abundances for these nuclei (Brachwitz et al. 2000; Maeda et al. 2010 -- but see also Seitenzahl et al. 2013).

\section{Sensitivity of nucleosynthesis to rate variations}

\subsection{Variation of all rates by a factor of 10}

\begin{figure}
\begin{center}  
\includegraphics[scale=0.8]{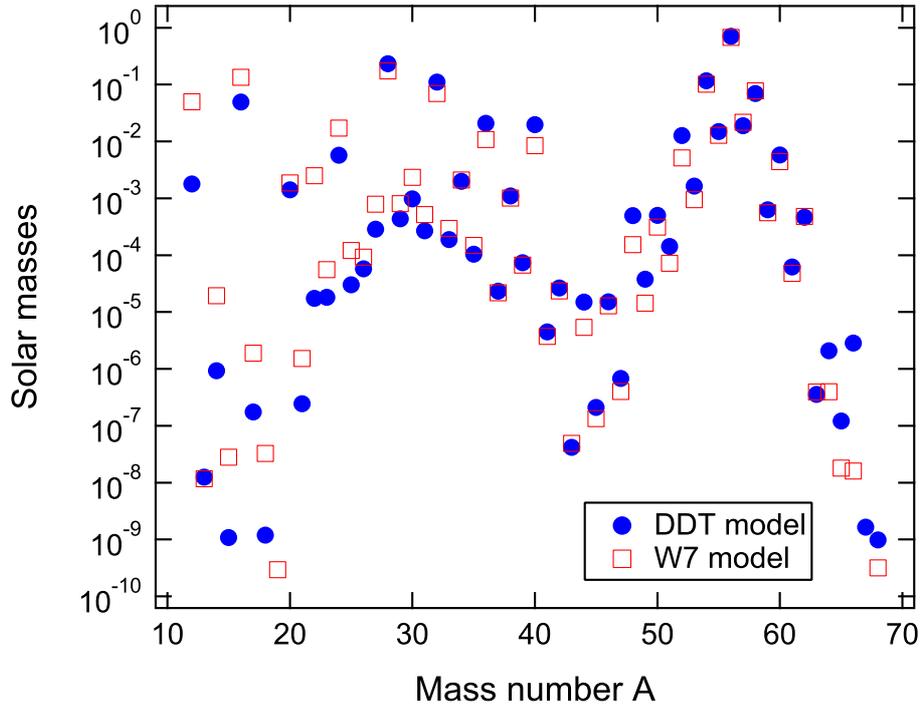}
\caption{(Colour online) Yields in M$_{\sun}$ vs. mass number A, as calculated for the DDT (filled circle) and W7 (empty square) models. These yields were obtained using standard rates in our nuclear network.}
\label{fig_std_yields}  
\end{center}  
\end{figure}

We have post-processed the thermodynamic histories from the delayed-detonation (DDT) and W7 thermonuclear supernova models described above.  This involved coupling an extended nuclear physics network to the temperature-density-time profiles extracted
from these two models.  Final yields were determined by summing the (mass-weighted) contributions from either all zones (for the W7 model) or all tracer particles (for the DDT model).  The nuclear physics network consisted of 443 species ranging from n to $^{86}$Kr (see Travaglio et al. 2004; Seitenzahl et al. 2009a).  Nuclear reactions such as ($p,\gamma$), ($\alpha,\gamma$), ($n,\gamma$), ($p,n$), ($\alpha,n$), ($\alpha,p$) have been included, as well as reactions such as $^{12}$C+$^{12}$C, $^{12}$C+$^{16}$O and $^{16}$O+$^{16}$O (along with all corresponding reverse processes).  Sufficient experimental information is available to determine rates for only a limited number of these reactions; these rates have been adopted from Iliadis et al. (2010), REACLIB V1.0 (Cyburt et al. 2010) and some recent updates for selected reactions.  Theoretical nuclear reaction rates, for the most part determined through Hauser-Feshbach models (e.g., Rauscher \& Thielemann 2000; Arnould \& Goriely 2003), were adopted when experimentally-based rates were not available.  For weak interactions, we have used stellar (temperature and density dependent) rates from the large-scale shell model calculations of Oda et al. (1994) and Langanke \& Mart\'inez-Pinedo (2000), supplemented with additional stellar rates from Pruet \& Fuller (2003) and Fuller et al. (1982).  Together, these sources provided stellar weak interaction rates for most species in our network with $A \geq 17$.  Nuclear statistical equilibrium was assumed above $T = 5$ GK.  The initial composition of the WD for both models was 47.5\% $^{12}$C, 50\% $^{16}$O and 2.5\% $^{22}$Ne, by mass.  Yields from the DDT and W7 models using our standard nuclear physics network are plotted in Fig.~\ref{fig_std_yields}.  For each of the two models, we then varied each individual rate in our standard network (together with the reverse process for reaction rates) by a factor of 10 (up and down), repeated the post-processing calculations for all thermodynamic histories, and compared the resulting yields with the yields shown in Fig.~\ref{fig_std_yields}.  

\begin{figure}  
\begin{center}  
\includegraphics[scale=0.8]{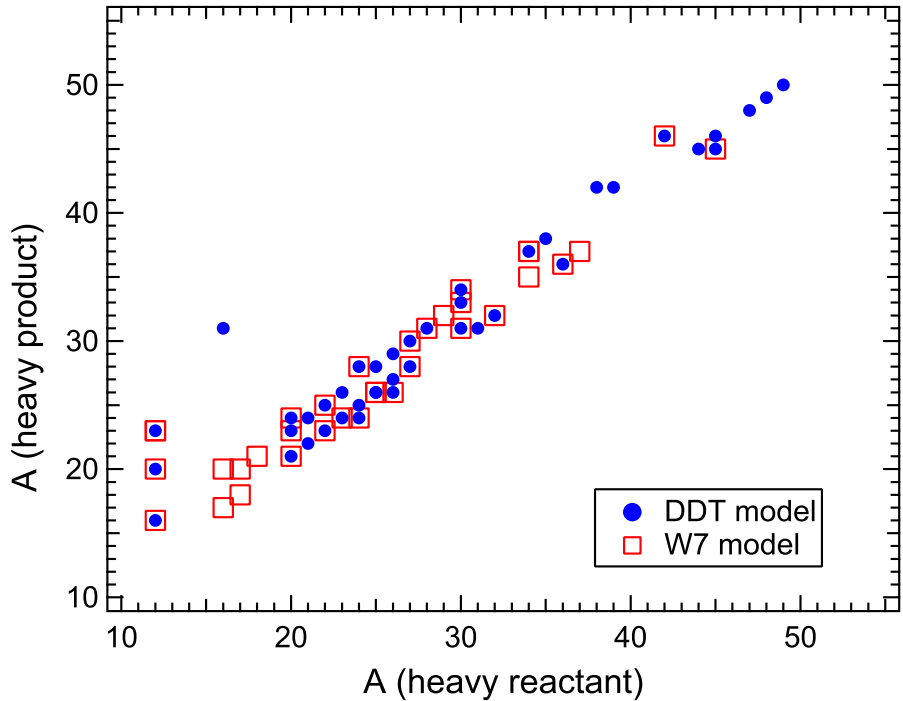}
\caption{(Colour online) Reactions whose rates affect the yield of at least one species in the DDT (filled circle) or W7 (empty square) models by at least a factor of two, when their rates are individually varied by a factor of 10.  The heavy product is plotted against the heavy reactant for all reactions in Table 1.  For example, the $^{22}$Ne($\alpha,n$)$^{25}$Mg reaction (variation of which affects yields in both models) would correspond to both an empty square and a filled circle at (x,y) = (22,25) here.}
\label{fig_important_reactions_both_models}  
\end{center}  
\end{figure}

\begin{figure}  
\begin{center}  
\includegraphics[scale=0.9]{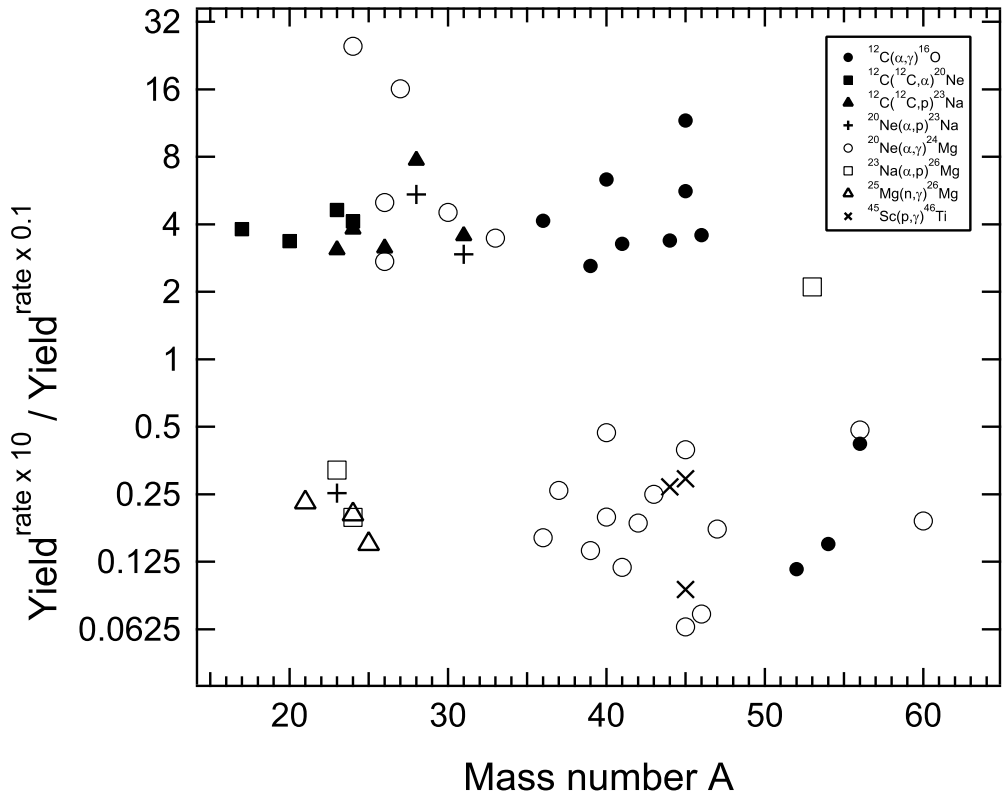}
\caption{Reaction rates whose variation by a factor of 10 affects the yields of at least three species in the DDT model (see Table 1).  For each reaction, we plot the yield of each affected nuclide when the rate was enhanced by a factor of 10 divided by the yield of that nuclide when the rate was reduced by a factor of 10.}
\label{fig_ddt_y10_y01}  
\end{center}  
\end{figure}

\begin{figure}  
\begin{center}  
\includegraphics[scale=0.9]{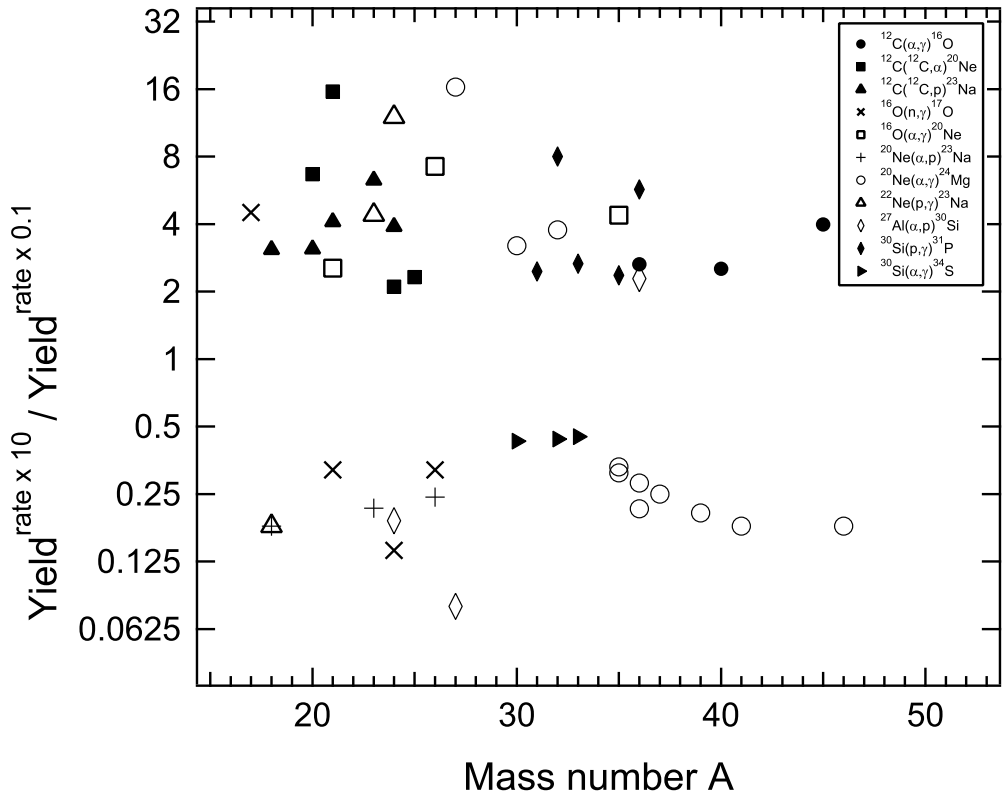}
\caption{Same as Fig.~\ref{fig_ddt_y10_y01} but for the W7 model.}
\label{fig_W7_y10_y01}  
\end{center}  
\end{figure}

For reaction rate variations, the results are summarized in Table 1.  Stable and radioactive species are listed since abundances were determined one hour after the beginning of the explosion, for each model.  To emphasize those rate variations that most significantly affected the yields in each model, we will restrict the discussion here to species that achieved an abundance of at least 10$^{-8}$ M$_{\sun}$ and deviated from the abundances determined with standard rates (i.e., Fig.~\ref{fig_std_yields}) by at least a factor of two (unless otherwise indicated).  We realize that for some applications the variation of yields of particular species by less than a factor of two may be of interest.  As such, full results on the specific effect of varying any rate in our network in either model are available upon request (but see also Section \ref{iso}).  Table 2 highlights important reactions from Table 1 by listing only those reaction rates that affected (i) the yields of at least \emph{three} species (again, all with abundances greater than 10$^{-8}$ M$_{\sun}$) by at least a factor of two in either the DDT or W7 models, and/or (ii) affected the yield of at least one species in \emph{both} models by at least a factor of two.  To facilitate interpretation, we also present the results of Table 1 in Figs.~\ref{fig_important_reactions_both_models}, \ref{fig_ddt_y10_y01} and \ref{fig_W7_y10_y01}.  For both the DDT and W7 models, Fig.~\ref{fig_important_reactions_both_models} shows the heavy product against the heavy reactant for all reactions in Table 1.  Figs.~\ref{fig_ddt_y10_y01} and \ref{fig_W7_y10_y01} show reaction rates from Table 1 whose variation affects the yields of at least three species in the DDT or W7 models, respectively.  For each reaction, these plots show a ratio: the yield of each affected species when the rate was enhanced by a factor of 10 divided by the yield of that species when the rate was reduced by a factor of 10.         

From the individual variation of nuclear reaction rates by a factor of 10, only 53 of the 2305 reactions in our network affect yields of any species with an abundance of at least 10$^{-8}$ M$_{\sun}$ by at least a factor of two, in either model.  (Note that for reactions, forward and reverse processes are not counted separately here since they must always be varied together.)  Of these, all but 9 reactions involve exclusively species with $A < 40$.  There are no reactions listed in Table 1 involving nuclei with $Z > 24$.  As can be seen from Table 2 (case C) and Fig.~\ref{fig_important_reactions_both_models}, 24 of these 53 reaction rates affect yields in both models.  The overall impact of these rate variations is limited however.  As seen in Table 2 (cases A and B) only 14 reactions affect the yields of three or more species, in either model.  Of these, the rates of the $^{12}$C($\alpha,\gamma$), $^{12}$C($^{12}C,\alpha$), $^{12}$C($^{12}C,p$), $^{20}$Ne($\alpha,p$), $^{20}$Ne($\alpha,\gamma$) and $^{30}$Si($p,\gamma$) reactions have the greatest impact, affecting the yields of at least five species when varied by a factor of 10; this is also illustrated in Figs.~\ref{fig_ddt_y10_y01} and \ref{fig_W7_y10_y01}.  Finally, only five reactions have an impact on the yields of the most abundant species produced (i.e., those with abundances greater than $10^{-2}$ M$_{\sun}$ in Fig.~\ref{fig_std_yields}): $^{12}$C($\alpha,\gamma$), $^{20}$Ne($\alpha,\gamma$), $^{24}$Mg($\alpha,\gamma$), $^{27}$Al($p,\gamma$), and $^{27}$Al($\alpha,p$).  Note that each of these five reactions also satisfies conditions for `case C' in Table 2. 

From the individual variation of weak interaction rates ($A > 17$) by a factor of 10, we first note that we have not varied the electron/positron capture rate and $\beta$-decay rate for a particular nucleus independently; rather, we only varied the sum of these two contributions to gain a measure of sensitivity to these processes.  Only three of the 658 weak interaction rates varied affect yields of species by at least 20\%, in either model: $^{28}$Si($\beta^{+}$)$^{28}$Al, $^{32}$S($\beta^{+}$)$^{32}$P, and $^{36}$Ar($\beta^{+}$)$^{36}$Cl.  From the enhancement of the stellar $^{28}$Si($\beta^{+}$) rate, the largest effects were seen in the yields of $^{48}$Ti, $^{50}$V, $^{55}$Mn, $^{57}$Fe (which increased by $\approx$40 -- 80\%) and $^{51}$V, $^{53}$Cr, $^{58}$Fe (which increased by a factor of $\approx$2).  Similarly, in the enhancement of the stellar $^{32}$S($\beta^{+}$) rate, the largest effects were seen in the yields of $^{44}$Ca, $^{50}$V, $^{52}$Cr, $^{54}$Mn, $^{56}$Fe, $^{57}$Fe, $^{59}$Co, $^{61}$Ni (which increased by $\approx$40 -- 80\%) and $^{48}$Ti, $^{51}$V, $^{53}$Cr, $^{55}$Mn, $^{58}$Fe (which increased by factors of $\approx$2 -- 3).  Yield changes were more modest from the multiplication of the $^{36}$Ar($\beta^{+}$) rate by a factor of 10: the yields of $^{48}$Ti, $^{51}$V, $^{53}$Cr, $^{55}$Mn, $^{58}$Fe all increased by $\approx$20\%.  No effects on yields were seen at levels greater than 20\% when any of these three rates was reduced by a factor of 10.  As well, no significant effect on yields was seen from the individual variation of any $\beta^{-}$ decay rate, in agreement with expectations (e.g., Brachwitz et al. 2000).

Finally, we note that since post-processing calculations only track existing thermodynamic histories, results obtained from variations in rates that significantly affect the energy production should be interpreted carefully. Indeed, a hydrodynamic code capable of suitably adjusting both the temperature and the density of the environment in response to any changes in energy generation is required to reliably treat such cases.  In the present calculations, only the $^{12}$C+$^{12}$C rates were observed to modify the overall energy output by more than 5\% at some point during the explosion in both models, when varied by a factor of 10.  As well, variations of the $^{12}$C+$^{16}$O and $^{16}$O+$^{16}$O rates affected the calculated energy generation in the DDT model only.  Of course, energy variation deduced from a post-processing study does not necessarily imply substantial energy variation in a hydrodynamic study.  Indeed, a recent hydrodynamic study (Bravo \& Mart\'inez-Pinedo 2012) found that variations of the $^{12}$C+$^{12}$C and $^{16}$O+$^{16}$O rates by a factor of 10 had a negligible impact on the energy of the supernova in their DDT model.

\subsection{Additional tests to motivate experiments}

We have examined the effects of varying thermonuclear reaction rates by a uniform factor of ten over all stellar temperatures.  To motivate and interpret experiments to better constrain the rates whose uncertainties have the largest impact on yields, we have performed three additional sets of post-processing calculations using the DDT and W7 models.  For all rates in Table 2 that affect the yields of at least three species in a model (11 rates for W7 and 8 rates for DDT), we have investigated the nucleosynthesis from (i) varying the rate by a uniform factor of two over all temperatures and (ii) varying the rate by a factor of two over four temperature windows: 0.01 -- 0.5 GK, 0.5 -- 1.0 GK, 1.0 -- 1.5 GK and 1.5 -- 2.0 GK.  To be clear, when a reaction rate (along with, as always, the rate of the reverse process) was varied within one of the four temperature windows, we retained the standard rate for all temperatures outside the chosen window.  These tests help to estimate particular temperature ranges over which the largest effects on yields may be expected from rate variations, so as to encourage laboratory measurements of e.g., reaction cross sections at the corresponding energies. Finally, (iii) we have examined the effect of using experimentally-based uncertainties (when available) for important reactions identified in Table 2.

When each member of this subset of reaction rates was individually varied by a factor of two over all temperatures, only the $^{12}$C($^{12}C,\alpha$) and $^{12}$C($^{12}C,p$) rates continued to significantly affect standard yields (by at least a factor of two, for species with standard abundances of at least 10$^{-8}$ M$_{\sun}$).  In the DDT model, the abundance of $^{28}$Mg was enhanced by a factor of $\approx2$ when the $^{12}$C($^{12}C,p$) rate was increased by a factor of two; in the W7 model, the abundances of $^{20}$Ne, $^{21}$Ne and $^{26}$Mg were enhanced by factors of $2-4$ when the $^{12}$C($^{12}C,\alpha$) rate was multiplied by a factor of two.

When each of these rates was individually varied by a factor of two within each of the four temperature windows mentioned above, only variations within the 1.5 -- 2.0 GK window significantly affected yields, for both models.  In the DDT model, variation of the rates of the $^{12}$C($^{12}C,\alpha$), $^{12}$C($^{12}C,p$), $^{20}$Ne($\alpha,p$), $^{23}$Na($\alpha,p$) and $^{25}$Mg($n,\gamma$) reactions by a factor of two within the 1.5 -- 2.0 GK window affected yields of species by at least a factor of two; in the W7 model, variation of the $^{12}$C($^{12}C,\alpha$), $^{12}$C($^{12}C,p$), $^{16}$O($n,\gamma$), $^{16}$O($\alpha,\gamma$), $^{20}$Ne($\alpha,p$) and $^{22}$Ne($p,\gamma$) rates within this window had an impact on yields.  Interestingly, these variations affected more isotopes than observed when varying the rates by a \emph{uniform} factor of two; moreover, for some species, the deviations from standard yields approached the level observed when varying rates by a uniform factor of 10.  For example, in the DDT model, enhancement of the $^{25}$Mg($n,\gamma$) rate by a uniform factor of 10 reduced the yields of $^{21}$Ne, $^{24}$Na and $^{25}$Mg by factors of 0.3 -- 0.4 (Table 1).   As mentioned above, enhancement of this rate by a uniform factor of two did not change the yield of any species by at least a factor of two.  However, enhancing this rate by a factor of two within the temperature window of 1.5 -- 2.0 GK reduced the yields of $^{21}$Ne, $^{24}$Na and $^{25}$Mg by factors of $0.4 - 0.5$.   These tests imply that one must clearly be cautious when using results from sensitivity studies to interpret experimental measurements where, e.g., a resonance may modify a given rate only over a limited range of temperatures.  As well, the lack of impact on yields from variations in the lower temperature windows suggests that useful measurements of these rates are probably best made at energies that correspond to temperatures above $1.5$ GK.

A final set of tests focused on the impact of using experimentally-based (temperature-dependent) uncertainties for important rates rather than variations by constant factors.  For the $^{16}$O($\alpha,\gamma$), $^{20}$Ne($\alpha,\gamma$), $^{22}$Ne($p,\gamma$), $^{30}$Si($p,\gamma$), and $^{20}$Ne($\alpha,p$) rates (all of which are in Table 2), we have calculated the nucleosynthesis in the DDT and W7 models assuming, for each rate, the ``high rate" and ``low rate" calculations of Iliadis et al. (2010).  For the $^{22}$Ne($\alpha,n$) rate, we have calculated the nucleosynthesis in the two models using the ``high" and ``low" rate calculations from Longland et al. (2012).  In general, between $\approx1 - 5$ GK, the high and low rates for each reaction differ by only $\approx 10 - 50\%$, although we note that the $^{22}$Ne($p,\gamma$), $^{22}$Ne($\alpha,n$) and $^{20}$Ne($\alpha,p$) rates are not well-constrained experimentally above 4 GK, 1.25 GK and 3.5 GK respectively.  At lower temperatures, differences can be as large as several orders of magnitude (for the $^{30}$Si($p,\gamma$) rate, from 0.01 -- 0.04 GK) but are usually within factors of $\approx2 - 40$ for temperatures between $\approx0.01 - 1$ GK.  We have also examined the effect of using two different calculations of the $^{12}$C($\alpha,\gamma$) rate (Kunz et al. 2002, and a re-evaluation of the Buchmann et al. 1996 rate from the REACLIB database (Cyburt et al. 2010)).  These two rates agree to within a factor of $\approx 2 - 3$ at relevant temperatures.  Note that we did not test the effects of using upper and lower limits for the $^{12}$C+$^{12}$C reactions as variation of these processes affects the nuclear energy generation rate in our models.  As such, post-processing studies cannot be used to examine such cases in detail (see section 3.1).  We found no significant effects (i.e., changes to any yields by a factor of at least two) when yields determined using the high rate for a reaction were compared to yields determined using the low rate for the same reaction, in either model.  As well, no significant difference was observed when comparing yields using the two different $^{12}$C($\alpha,\gamma$) rates.  This is consistent with the results from the tests using the temperature windows: although uncertainties at low temperatures may be comparable to or even  larger than a factor of 10, these have little impact on the yields.  For these cases then, we encourage experimentalists to confirm the nuclear physics input presently used to determine these rates for $T \gtrsim 1.5$ GK, or, for the $^{22}$Ne($p,\gamma$), $^{22}$Ne($\alpha,n$) and $^{20}$Ne($\alpha,p$) reactions, perform measurements to better determine these rates up to $\approx5$ GK (at which NSE was assumed in our calculations).

\subsection{Impact on predicted isotopic ratios}
\label{iso}         
In principle, abundance ratios of radioactive species produced in SN~Ia may be inferred from late time bolometric light curves. The longest lived members of the contributing nuclear decay chains (starting from $^{44}$Ti, $^{55}$Co, $^{56}$Ni, and $^{57}$Ni) have half-lives that are well separated.  This leads to ``ankles" in the lepton-dominated bolometric light curves when a longer lived chain becomes the dominant heat source (Seitenzahl et al. 2009b; Seitenzahl 2011). If the production factors are sufficiently different, this effect can in turn be used to distinguish between competing explosion models (R\"opke et al. 2012).  In the two models examined here, the net production of $^{44}$Ti is dominated by the rates of the $^{44}$Ti($\alpha,p$), $^{40}$Ca($\alpha,\gamma$), $^{44}$Ti($p,\gamma$), $^{44}$Ti($n,p$) and $^{44}$Ti($n,\gamma$) reactions.  The net production of $^{55}$Co is dominated by the rates of the $^{55}$Co($n,p$), $^{54}$Fe($p,\gamma$), $^{55}$Co($p,\gamma$), $^{55}$Co($n,\gamma$) and $^{55}$Co($p,n$) reactions.  Finally, the $^{56}$Ni($n,p$), $^{56}$Ni($n,\gamma$), $^{56}$Ni($p,\gamma$) and $^{55}$Co($p,\gamma$) reaction rates are primarily responsible for the net production of $^{56}$Ni, while the $^{56}$Ni($n,\gamma$), $^{57}$Ni($n,p$), $^{56}$Co($p,\gamma$) and $^{57}$Ni($p,\gamma$) reactions produce $^{57}$Ni.

It is of interest to test to what extent model predictions of these ratios are sensitive to the input reaction and weak interaction rates.  While this could be inferred from the information in Table 1, the restrictions we have placed upon the contents of that table for the sake of brevity (i.e., changes of abundances with respect to standard yields by at least a factor of two) suggest that a separate analysis be used to account for changes in both members of a particular abundance ratio.  We have examined the effects of individual rate variations by a (uniform) factor of 10 on ratios composed of the species $^{44}$Ti, $^{55}$Co, $^{56}$Ni and $^{57}$Ni.  In the DDT model, we find that variations in the $^{12}$C($\alpha,\gamma$) rate affects the ratios $^{44}$Ti/$^{55}$Co and $^{44}$Ti/$^{56}$Ni by a factor of $\approx$2; individual variation of the $^{20}$Ne($\alpha,\gamma$) and $^{44}$Ti($\alpha,p$) rates affect the same ratios by $\approx$30\%.  As well, variation of the $^{55}$Co($p,\gamma$) rate modifies the $^{44}$Ti/$^{55}$Co, $^{44}$Ti/$^{56}$Ni, $^{55}$Co/$^{56}$Ni and $^{55}$Co/$^{57}$Ni ratios by $\approx$30\%.  In the W7 model, variation of the $^{12}$C($\alpha,\gamma$) and $^{40}$Ca($\alpha,\gamma$) rates affects the ratios of $^{44}$Ti/$^{55}$Co and $^{44}$Ti/$^{56}$Ni by $\approx$30\%.  The $^{44}$Ti($\alpha,p$) and $^{55}$Co($p,\gamma$) rates have somewhat larger influence when varied in W7; both change the ratios $^{44}$Ti/$^{55}$Co and $^{44}$Ti/$^{56}$Ni by $\approx40\%$, while the latter also modifies the $^{55}$Co/$^{56}$Ni and $^{55}$Co/$^{57}$Ni ratios at about the same level.  No weak interaction rates have comparable effects on ratios of these radioactive species, in either model, when individually varied by a factor of 10.  

The rates affecting the ratios of concern include those rates directly responsible for the net production of the relevant isotopes, as mentioned above (e.g., $^{44}$Ti($\alpha,p$), $^{55}$Co($p,\gamma$), and $^{40}$Ca($\alpha,\gamma$)) as well as other rates such as $^{12}$C($\alpha,\gamma$) and $^{20}$Ne($\alpha,\gamma$) which affect the abundances of numerous species when varied by a factor of 10 (see Table 1).  As well, we find that most of the ratios affected by rate variations include $^{44}$Ti.  This is because the abundances of $^{55}$Co, $^{56}$Ni and $^{57}$Ni are not very sensitive to changes of rates by a factor of 10, even when one considers abundance changes below a factor of two.  The abundances of $^{56}$Ni and $^{57}$Ni are robust at the 10\% level to all rate changes; the abundance of $^{55}$Co changes by $\approx30\%$ due to variation of the $^{55}$Co($p,\gamma$) rate.  We conclude by noting that for the two models tested here, the predicted ratios of radionuclides are rather robust to rate uncertainties.  Given that delayed-detonation and merger scenarios predict ratios of e.g., $^{55}$Co/$^{56}$Ni that differ by a factor of $\approx3$  (R\"opke et al. 2012), our results further support the idea of using observed ratios of radioactive species as a means of discriminating between explosion mechanisms.

\section{Discussion}  

Of the 2305 nuclear reactions in our network, individual variation of only 29 rates by a factor of 10 significantly affects yields of either at least three species in one model, or at least one species in both SN~Ia models examined in this study (Table 2).  Moreover, all but eight of these 29 reactions exclusively involve species with $20 \leq A \leq 36$ and $10 \leq Z \leq 17$ (that is, species outside the Fe group).  No reaction involving any species with $Z > 24$ (Table 1) was found to affect yields in either model by at least a factor of two (for species with standard yields greater than 10$^{-8}$ M$_{\sun}$).  Variations in only three reaction rates (those of $^{12}$C($\alpha,\gamma$), $^{20}$Ne($\alpha,\gamma$) and $^{23}$Na($\alpha,p$)) affect the yields of Fe-group species ($Z > 23$, see Table 1). Reactions that have the greatest influence on supernova nucleosynthesis in our models (when their rates are varied by a factor of 10) include $^{12}$C($\alpha,\gamma$), $^{12}$C+$^{12}$C, $^{20}$Ne($\alpha,p$), $^{20}$Ne($\alpha,\gamma$) and $^{30}$Si($p,\gamma$), all of which have a significant impact on the yields of at least five species in a model.    Additional tests where the most influential rates were varied only within specific temperature windows indicate that rate variations for $T > 1.5$ GK have more impact on yields than variations at lower temperatures.  These results are broadly consistent with the work of Bravo \& Mart\'inez-Pinedo (2012) in which thermodynamic histories from a single 1D delayed-detonation model were post-processed to determine nuclear reactions whose rates influenced SN~Ia yields when varied by a factor of ten.  (The impact of variations in weak interaction rates was not examined in that study.)  They found that nucleosynthesis in their model was most sensitive to the $^{20}$Ne($\alpha,\gamma$), $^{24}$Mg($\alpha,p$) and $^{30}$Si($p,\gamma$) rates, and that reactions involving species with $Z > 22$ did not have considerable impact on yields when their rates were varied by a factor of 10. As well, they allude to the relative robustness of the yields of Fe peak nuclei and claim that rate modifications have largest impact between about $2 < T < 4$ GK.  This general agreement between the results from independent studies involving three different underlying models is encouraging, and suggests that SN~Ia yields in single degenerate scenarios with Chandrasekhar-mass WDs are not very sensitive to uncertainties in nuclear reaction rates.  This is further supported by how most of the influential reactions mentioned above (e.g., $^{12}$C($\alpha,\gamma$), $^{20}$Ne($\alpha,p$), $^{20}$Ne($\alpha,\gamma$), $^{30}$Si($p,\gamma$)) have thermonuclear rates that are at least partially based upon experimental measurements, and which have uncertainties that do not significantly affect our calculated yields (see Section 3.2).  Given this, as well as the increasing reliability of statistical model methods with increasing stellar temperature, it is likely that the impact of even the limited set of reactions in Table 2 is overestimated when a variation factor of 10 is employed.  Nonetheless, as stated in section 3.2, we found that even rate variations by a factor of two within the $1.5 < T < 2.0$ GK temperature window affected yields in our models.  

Of the 658 weak interaction rates varied within our network, only three stellar rates (where contributions from electron/positron capture and $\beta$-decay were summed) have even a modest impact on yields when individually varied by a factor of 10, and then only when enhanced.  This is interesting in light of the study by Brachwitz et al. (2000) who examined the role of different libraries of electron capture rates on SN~Ia yields.  In general, they found that the use of libraries with larger electron capture rates resulted in larger yields (by factors of $\approx 2 - 10$ or more) of neutron-rich species in the Fe group (e.g., $^{50}$Ti, $^{54}$Cr, $^{58}$Fe, $^{64}$Ni).  Note that the different libraries adopted in that study differed by more than simple uniform scalings of all rates, and that the largest electron capture rate differences between libraries (up to several orders of magnitude) occurred for odd-A parent nuclei.  In the present study, we did find some effect on neutron-rich nuclei (e.g., $^{57,58}$Fe, $^{53}$Cr, see section 3.1) from the enhancement of individual rates.  To investigate this issue further, we examined the effect on yields from multiplying all weak interaction rates in our network (simultaneously) by a uniform factor of 10, in the W7 model.  This is obviously not equivalent to the study of Brachwitz et al. (2000) since we use different underlying models, different rate libraries, and we enhance the total weak interaction rates for all species ($\beta$-decay and electron capture) as opposed to only the electron capture contributions; nonetheless, we can test the general trend observed.  Indeed, we found significant overproduction of neutron-rich species between $A \approx 49 - 65$ when all weak rates were enhanced.  For example, $^{50}$Ti, $^{54}$Cr, $^{58}$Fe, and $^{64}$Ni (highlighted in Brachwitz et al. 2000) were overproduced by factors of $\approx$ 30, 12, 4 and 8 respectively.  As well, $^{49}$Sc, $^{49}$Ti, $^{51,52}$V, $^{56}$Mn, $^{59}$Fe, $^{60, 61}$Co, and $^{63}$Ni (all neutron rich) exhibited large overproduction factors of $\approx 5 - 10$ relative to yields with the standard rates.  At least with regard to weak interaction rates involved in SN~Ia nucleosynthesis, then, these results both expose limitations of the individual variation method for sensitivity studies and confirm the need for a detailed, consistent treatment of all relevant rates.

\section{Conclusions}

We have investigated the sensitivity of yields from two different thermonuclear supernova models to variations in nuclear reaction and weak interaction rates.  Thermodynamic histories from a delayed-detonation model and the canonical W7 pure deflagaration model were post-processed in conjunction with the individual variation of each rate in our network by a uniform factor of ten.  The rates of only fourteen reactions significantly affect the yields of at least three species in a model by a factor of two, for species with standard yields greater than 10$^{-8} \mathrm{M_{\sun}}$.  Of these, the rates of the $^{12}$C($\alpha,\gamma$), $^{12}$C+$^{12}$C, $^{20}$Ne($\alpha,p$), $^{20}$Ne($\alpha,\gamma$) and $^{30}$Si($p,\gamma$) reactions had the greatest impact on nucleosynthesis.  Weak interaction rates had a relatively more modest impact on yields when individually varied.  Enhancement of the stellar $^{28}$Si($\beta^{+}$)$^{28}$Al, $^{32}$S($\beta^{+}$)$^{32}$P and $^{36}$Ar($\beta^{+}$)$^{36}$Cl rates affected some yields by a factor of $\approx 2$; on the other hand, no significant effect on yields was noted in either model when any weak interaction rate was reduced by a factor of ten.   In general, rates that had an impact on the calculated nucleosynthesis involved nuclei with $Z \leq 24$ and $A \lesssim 40$; variation of these rates mostly affected nuclei with $20 \lesssim A \lesssim 45$, with the abundances of Fe-group nuclei being rather robust to rate variations.  This is likely due to how, in models involving Chandrasekhar-mass WDs, most species in the Fe-group are synthesized in NSE (which is insensitive to rate variations).  The abundances of Fe-group nuclei may indeed be less robust to rate variations in a merger scenario, where a larger fraction of the material in the Fe-group is synthesized in incomplete Si-burning.  Additional tests involved the variation of important nuclear reaction rates by a factor of two (over all temperatures), by a factor of two only within specific temperature windows, and by experimentally-based uncertainties.  In the first of these additional tests, only the $^{12}$C+$^{12}$C rates continued to affect yields; in the second, it was found that variations for $T < 1.5$ GK did not have a large impact on yields in our models; in the third test, it was observed that for the reactions examined, experimentally-based uncertainties did not result in significant changes to calculated yields, in either model.  

Overall, given the size of the reaction network employed (443 species from H to $^{86}$Kr), nucleosynthesis in our two adopted models involving single degenerate scenarios with Chandrasekhar-mass WDs is rather robust to individual variations of the input rates.  Laboratory measurements of the few important nuclear reaction rates (see Table 2), especially at energies corresponding to temperatures above 1.5 GK, would be welcome to further constrain the model predictions.  Nonetheless, our results support the idea of using isotopic ratios of radioactive species as a means of discerning between single and double degenerate scenarios (which may differ by factors of up to $\approx$3 in their predictions of these ratios).   

Finally, we stress the need for some caution in interpreting our results.  We have examined the role of rate variations in delayed-detonation and pure deflagaration models with Chandrasekhar-mass WDs.  Given the scale of the required calculations, we have limited the scope of this work to investigating to what degree nucleosynthesis predictions from these different, representative, explosion simulations are affected by varying the input nuclear reaction and weak interaction rates. We have not examined the role of rate variations in other proposed SNe~Ia scenarios, such as those involving mergers or sub-Chandrasekhar-mass WDs, nor have we tested the role of rate variations in determining, e.g., the early evolution of the progenitor in our models.  As well, we have not explored the impact of the initial composition of the white dwarf.  Such studies would be valuable and are encouraged.  Moreover, we have individually varied all rates, but have not examined in detail the effects of simultaneous variations of rates.  In other astrophysical scenarios, sensitivity studies using individual variation and simultaneous variation methods gave similar results (e.g., for Type I X-ray bursts, see Parikh et al. 2008).  This should be confirmed for thermonuclear supernovae, especially given our observation of large abundance changes when all weak interaction rates in our network were simultaneously enhanced by a factor of 10.  As such, it is clear that a consistent set of stellar weak interaction rates for all nuclei involved in models of Type Ia supernovae is urgently needed.  

\begin{acknowledgements}  
We thank K. Maeda and K. Nomoto for providing the thermodynamic histories of the zones of the W7 explosion model.
This  work has  been partially supported  by the Spanish MICINN grants
AYA2010-15685 and EUI2009-04167, the Government of Catalonia grant 2009SGR-1002, the E.U. FEDER funds, and
the ESF EUROCORES Program EuroGENESIS.  The simulations presented here were carried out in part at
the Computer Center of the Max Planck Society, Garching,
Germany. IRS and FKR were partially supported by the Deutsche
Forschungsgemeinschaft through the graduate school ``Theoretical
Astrophysics and Particle Physics" at the University
of W\"urzburg (GRK 1147). FKR received further support from the Deutsche Forschungsgemeinschaft through the Emmy Noether Program (RO 3676/1-1) and from the ARCHES prize of the German Federal Ministry of Education and Research (BMBF).
\end{acknowledgements}


\onecolumn
\clearpage


\onecolumn

\begin{center}

\begin{longtable}[ht]{|l|l|l l|l l|}
  \caption{Nuclear reaction rates that significantly affect abundances in the DDT and W7 models when varied by a factor of 10 (up or down).  Yields are presented relative to those using standard rates (i.e., Fig.~\ref{fig_std_yields}).  Only species that achieve an abundance of at least 10$^{-8}$ M$_{\sun}$ and deviate from the standard abundances by at least a factor of two are presented here.} \\

\hline
    Reaction&Nuclide&\multicolumn{2}{c|}{W7 model}&\multicolumn{2}{c|}{DDT model}\\ \hline
    &&x10&x0.1&x10&x0.1\\ \hline
   \endfirsthead

\multicolumn{6}{c}{{\tablename\ \thetable{} -- continued from previous page}} \\ \hline
Reaction&Nuclide&\multicolumn{2}{c|}{W7 model}&\multicolumn{2}{c|}{DDT model}\\ \hline
&&x10&x0.1&x10&x0.1\\ \hline
\endhead

\hline \multicolumn{6}{|r|}{{Continued on next page...}} \\ \hline
\endfoot

\hline 
\endlastfoot

   $^{12}$C($\alpha,\gamma$)$^{16}$O& $^{36}$Ar& 2.2 &  & 3.0 & \\
                        & $^{39}$K&&&2.0&\\
                        & $^{40}$Ca&2.2&&4.6&\\
                        & $^{41}$Ca&&&2.4&\\
                        & $^{44}$Ti&&&2.4&\\
                        & $^{45}$Ti&3.2&&7.1&\\
                        & $^{45}$Sc&&&4.3&\\
                        & $^{46}$Ti&&&2.5&\\
                        & $^{52}$Cr&&&0.17&\\
                        & $^{54}$Mn&&&0.21&\\
                        & $^{56}$Fe&&&0.42&\\

   $^{12}$C($^{12}$C,$\alpha$)$^{20}$Ne& $^{17}$O&  &  &2.6 &\\
                        & $^{20}$Ne& 5.0 &  &2.3 &\\
                        & $^{21}$Ne&11&&&\\
                       & $^{23}$Na&  &  &3.7 &\\
                        & $^{24}$Na&2.5&&3.3&\\
                        & $^{25}$Mg&2.1&&&\\
                        & $^{26}$Mg&5.2&&&\\

   $^{12}$C($^{12}$C,$p$)$^{23}$Na& $^{18}$O& 2.1 &  & &\\
                        & $^{20}$Ne&3.7&&&\\
                        & $^{21}$Ne&6.7&&&\\
                        & $^{23}$Na&2.6&0.41&&0.45\\
                        & $^{24}$Na&2.7&&3.1&\\
                        & $^{26}$Mg&4.0&&&\\
                        & $^{26}$Al&2.1&&2.5&\\
                        & $^{28}$Mg&&&6.2&\\
                        & $^{31}$Si&&&2.8&\\

   $^{12}$C($^{12}$C,$n$)$^{23}$Mg& $^{21}$Ne& 2.9 &  &&\\
                        & $^{26}$Mg&2.0&&&\\
                     
   $^{16}$O($n,\gamma$)$^{17}$O& $^{17}$O& &0.31  && \\
                        & $^{21}$Ne&& 2.8&&\\
                        & $^{24}$Na&&4.0&&\\
                        & $^{26}$Mg&&2.7&&\\

   $^{16}$O($\alpha,\gamma$)$^{20}$Ne& $^{21}$Ne& 2.5&  &&\\
                        & $^{26}$Al&6.4&&&\\
                        & $^{35}$S&3.9&&&\\

 $^{16}$O($^{16}$O,$p$)$^{31}$P& $^{45}$Ti& &  & 2.4 &\\

   $^{17}$O($p,\gamma$)$^{18}$F& $^{18}$O& 8.2& 0.21 & & \\
                       
  $^{17}$O($\alpha,n$)$^{20}$Ne& $^{17}$O&0.35 &  & &\\

   $^{18}$O($\alpha,n$)$^{21}$Ne& $^{18}$O& 0.32& & &\\
   $^{20}$Ne($n,\gamma$)$^{21}$Ne& $^{21}$Ne& 6.2& 0.24 & 4.6 & 0.16\\

   $^{20}$Ne($\alpha,p$)$^{23}$Na& $^{18}$O& 0.44& 2.4 &  &\\
                        & $^{23}$Na&0.47& 2.2&0.48&\\
                        & $^{26}$Al&&2.1&&\\
                        & $^{28}$Mg&& &5.4&\\
                        & $^{31}$Si&&&3.0&\\

   $^{20}$Ne($\alpha,\gamma$)$^{24}$Mg& $^{24}$Mg& 3.1 & 0.07 & 5.1 & 0.21\\
                        & $^{26}$Mg&&&2.1&\\
                        & $^{26}$Al&&&2.9&\\
                        & $^{27}$Al&&0.12&3.4&0.21\\
                        & $^{30}$Si&&0.36&&0.42\\
                        & $^{32}$P&2.2&&&\\
                        & $^{33}$P&&&&0.49\\
                        & $^{35}$S&&3.0&&\\
                        & $^{35}$Cl&&2.2&&\\
                        & $^{36}$Cl&0.38&&&\\
                        & $^{36}$Ar&0.36&&0.29&\\
                        & $^{37}$Ar&0.45&&0.39&\\
                        & $^{39}$K&0.32&&0.25&\\
                        & $^{40}$K&&&&2.1\\
                        & $^{40}$Ca&&&0.41&2.1\\
                        & $^{41}$Ca&0.29&&0.22&\\
                        & $^{42}$Ca&&&0.32&\\
                        & $^{43}$Ca&&&0.44&\\
                        & $^{45}$Sc&&&&2.5\\
                        & $^{45}$Ti&&&0.19&3.0\\
                        & $^{46}$Ti&0.24&&0.15&2.1\\
                        & $^{47}$Ti&&&0.29&\\
                        & $^{56}$Fe&&&0.43&\\
                        & $^{60}$Ni&&&0.24&\\

   $^{21}$Ne($n,\gamma$)$^{22}$Ne& $^{21}$Ne& &  & 0.44 &\\

   $^{21}$Ne($\alpha,n$)$^{24}$Mg& $^{21}$Ne& &  & 0.17 & 3.1\\

   $^{22}$Ne($p,\gamma$)$^{23}$Na& $^{18}$O& 0.42& 2.3 && \\
                        & $^{23}$Na&2.3& &&\\
                        & $^{24}$Na&4.0&0.33&2.2&\\

   $^{22}$Ne($\alpha,n$)$^{25}$Mg& $^{17}$O& & 0.36 & &\\
                        & $^{18}$O&2.5&0.23 &&\\
                      & $^{24}$Na& & & & 0.44\\

   $^{23}$Na($n,\gamma$)$^{24}$Na& $^{24}$Na&8.1 & 0.10 & 4.0 & 0.14\\

   $^{23}$Na($\alpha,p$)$^{26}$Mg& $^{23}$Na& &  &0.47 &\\
                        & $^{24}$Na&&&0.30&\\
                      & $^{53}$Cr& & & 2.1& \\

   $^{24}$Na($p,n$)$^{24}$Mg& $^{24}$Na& 0.16&  &0.13 & 2.8\\
   $^{24}$Mg($n,\gamma$)$^{25}$Mg& $^{26}$Al& &  &3.9 & 0.38\\
   $^{24}$Mg($\alpha,\gamma$)$^{28}$Si& $^{24}$Mg& 0.31&  & 0.34 &\\
                        & $^{27}$Al&0.42& &0.34&\\

   $^{25}$Mg($n,\gamma$)$^{26}$Mg& $^{21}$Ne& &  &0.41 &\\
                       & $^{24}$Na& &  & 0.35 & \\
                       & $^{25}$Mg& &  & 0.30 &\\
                       & $^{26}$Mg& 4.3&  0.45&&\\
                        
   $^{25}$Mg($p,\gamma$)$^{26}$Al& $^{26}$Al& 6.2&  & 2.2 &\\

   $^{25}$Mg($\alpha,n$)$^{28}$Si& $^{26}$Al& &  &0.34 &2.1\\
                       & $^{31}$Si& &  & 2.2 & \\

   $^{26}$Mg($n,\gamma$)$^{27}$Mg& $^{28}$Mg& &  & 5.4&\\
                       & $^{31}$Si& &  & 2.2 & \\

   $^{26}$Mg($p,n$)$^{26}$Al& $^{26}$Al& 0.40& & 0.32& \\

   $^{26}$Mg($\alpha,n$)$^{29}$Si& $^{26}$Mg& &  &0.44 &\\
                       & $^{31}$Si& &  & 3.0 & \\

   $^{27}$Al($p,\gamma$)$^{28}$Si& $^{24}$Mg& 0.28&  & 0.34 & \\
                        & $^{27}$Al&0.26& &0.20&\\

   $^{27}$Al($\alpha,p$)$^{30}$Si& $^{24}$Mg& 0.25& & 0.32 &\\
                        & $^{27}$Al&0.16&2.0 &0.14&2.1\\
                        & $^{36}$S&2.1&&&\\

   $^{28}$Si($\alpha,p$)$^{31}$P& $^{31}$P& & 2.1& & \\
                        & $^{36}$S&&0.31 &&0.40\\

   $^{29}$Si($\alpha,n$)$^{32}$S& $^{29}$Si&2.1&  &  & \\

   $^{30}$Si($p,\gamma$)$^{31}$P& $^{31}$P& & 0.49& & \\
                        & $^{32}$P&2.0&0.25 &&\\
                        & $^{33}$P&&0.48&&\\
                        & $^{35}$S&2.4& &&\\
                        & $^{36}$S&&0.33&&\\

   $^{30}$Si($\alpha,\gamma$)$^{34}$S& $^{30}$Si&0.48 & & & \\
                        & $^{32}$P&0.49&&&\\
                        & $^{33}$P& 0.50&&0.50&\\
   $^{30}$Si($n,\gamma$)$^{31}$Si& $^{31}$Si& & &5.1 &\\
   $^{30}$Si($\alpha,n$)$^{33}$S& $^{33}$P&0.48 & 3.1& & 3.0\\
                        & $^{36}$S&&2.6&&2.8\\
   $^{31}$Si($p,n$)$^{31}$P& $^{31}$Si&& & & 3.9\\

 $^{32}$P($p,n$)$^{32}$S& $^{32}$P& 0.42&  & 0.48 & \\

 $^{34}$S($p,\gamma$)$^{35}$Cl& $^{35}$S& &2.1  & & \\

 $^{34}$S($\alpha,p$)$^{37}$Cl& $^{37}$Cl& 0.29&  & 0.43& \\

 $^{34}$S($\alpha,n$)$^{37}$Ar& $^{37}$Cl& 0.37&  && \\

 $^{35}$Cl($\alpha,p$)$^{38}$Ar& $^{44}$Ca& &  & 2.6 &\\

 $^{36}$S($p,n$)$^{36}$Cl& $^{36}$S& &0.35  & & 0.44\\

 $^{37}$Cl($p,n$)$^{37}$Ar& $^{37}$Cl& &0.44  & &\\

 $^{38}$Ar($\alpha,\gamma$)$^{42}$Ca& $^{44}$Ca& &  & 2.2& \\
 $^{39}$K($\alpha,p$)$^{42}$Ca& $^{44}$Ca& &  & 2.5& \\
   $^{42}$Ca($\alpha,\gamma$)$^{46}$Ti& $^{38}$Ar&0.50 & & & \\
                        & $^{42}$Ca&0.42&&&\\
                        & $^{46}$Ti&&&2.2&\\
                        & $^{47}$Ti&&&2.1&\\
   $^{44}$Ca($p,\gamma$)$^{45}$Sc& $^{44}$Ca&& & &2.7 \\

   $^{45}$Sc($p,\gamma$)$^{46}$Ti& $^{44}$Ca&& & &3.7 \\
                        & $^{45}$Sc&&&0.41&4.4\\
                        & $^{45}$Ti&&&&2.4\\

 $^{45}$Sc($p,n$)$^{45}$Ti& $^{45}$Ti& &2.1  & 0.43 & 3.0\\

   $^{47}$Ti($n,\gamma$)$^{48}$Ti& $^{47}$Ti&& & 0.35& \\
   $^{48}$Ti($p,\gamma$)$^{49}$V& $^{48}$Ti&& & &2.4 \\

   $^{49}$V($p,\gamma$)$^{50}$Cr& $^{48}$Ti&& & &4.4 

\label{table_variations}  
\end{longtable}
\end{center}


\begin{table}[ht]
  \caption{Summary of important reactions from Table 1.  Variation of each of the listed rates by a factor of 10 (up or down) modified yields of at least \emph{three} species by at least a factor of two in the W7 model (``Case A") or the DDT model (``Case B").  If variation of the rate affected the yield of at least one species in \emph{both} models, it is designated as ``Case C".}
  \begin{center}
    \begin{tabular}{|l|l|l|l|}

    \hline
    Reaction&\multicolumn{3}{c|}{Importance}\\
   \hline
    &Case A&Case B&Case C\\
   \hline
$^{12}$C($\alpha,\gamma$)$^{16}$O&X&X&X\\
$^{12}$C($^{12}$C,$\alpha$)$^{20}$Ne&X&X&X\\
$^{12}$C($^{12}$C,$p$)$^{23}$Na&X&X&X\\
$^{16}$O($n,\gamma$)$^{17}$O&X&&\\
$^{16}$O($\alpha,\gamma$)$^{20}$Ne&X&&\\
$^{20}$Ne($n,\gamma$)$^{21}$Ne&&&X\\
$^{20}$Ne($\alpha,p$)$^{23}$Na&X&X&X\\
$^{20}$Ne($\alpha,\gamma$)$^{24}$Mg&X&X&X\\
$^{22}$Ne($p,\gamma$)$^{23}$Na&X&&X\\
$^{22}$Ne($\alpha,n$)$^{25}$Mg&&&X\\
$^{23}$Na($n,\gamma$)$^{24}$Na&&&X\\
$^{23}$Na($\alpha,p$)$^{26}$Mg&&X&\\
$^{24}$Na($p,n$)$^{24}$Mg&&&X\\
$^{24}$Mg($\alpha,\gamma$)$^{28}$Si&&&X\\
$^{25}$Mg($n,\gamma$)$^{26}$Mg&&X&X\\
$^{25}$Mg($p,\gamma$)$^{26}$Al&&&X\\
$^{26}$Mg($p,n$)$^{26}$Al&&&X\\
$^{27}$Al($p,\gamma$)$^{28}$Si&&&X\\
$^{27}$Al($\alpha,p$)$^{30}$Si&X&&X\\
$^{28}$Si($\alpha,p$)$^{31}$P&&&X\\
$^{30}$Si($p,\gamma$)$^{31}$P&X&&\\
$^{30}$Si($\alpha,\gamma$)$^{34}$S&X&&X\\
$^{30}$Si($\alpha,n$)$^{33}$S&&&X\\
$^{32}$P($p,n$)$^{32}$S&&&X\\
$^{34}$S($\alpha,p$)$^{37}$Cl&&&X\\
$^{36}$S($p,n$)$^{36}$Cl&&&X\\
$^{42}$Ca($\alpha,\gamma$)$^{46}$Ti&&&X\\
$^{45}$Sc($p,\gamma$)$^{46}$Ti&&X&\\
$^{45}$Sc($p,n$)$^{45}$Ti&&&X\\

   \hline
  \end{tabular}
  \end{center}
\label{summary}  
\end{table}

\end{document}